\documentclass[aps,prd,a4paper,twocolumn,floatfix,showpacs,nofootinbib]{revtex4-1}
\usepackage{amsmath,amsfonts,amssymb,mathrsfs}
\usepackage{graphicx}
\usepackage{subfigure}
\usepackage{enumerate}
\usepackage{color}
\def\be{\begin{equation}}
\def\ee{\end{equation}}

\def\beq{\begin{equation}}
\def\eneq{\end{equation}}
\def\bea{\begin{eqnarray}}
\def\enea{\end{eqnarray}}

\def\bea{\begin{eqnarray}}
\def\eea{\end{eqnarray}}

\begin{document}
\title{Ho\v rava--Lifshitz gravity: detailed balance revisited}

\author{Daniele Vernieri and Thomas P.\ Sotiriou}
\affiliation{SISSA--ISAS, Via Bonomea 265, 34136, Trieste, 
Italy {\rm and} INFN, Sezione di Trieste, Italy}
\begin{abstract}
We attempt a critical reconsideration of ``detailed balance'' as a principle that can be used to restrict the proliferation of couplings in Ho\v rava--Lifshitz gravity. We re-examine the shortcomings that have been usually associated with it in the literature and we argue that easy remedies can be found for all of them  within the framework of detailed balance, and that the most persistent of them are actually related to projectability. We show that, once projectability is abandoned, detailed balance reduces the number of independent couplings by roughly an order of magnitude and imposes only one restriction that constitutes a phenomenological concern: the size of the (bare) cosmological constant is unacceptably  large. Remarkably, this restriction (which is present in the projectable version as well) has been so far under-appreciated in the literature. Optimists might prefer to interpret it as a potential blessing in disguise, as it allows one to entertain the idea of a miraculous cancellation between the bare cosmological constant and the (still poorly understood) vacuum energy contribution.
\end{abstract}
\pacs{04.60.-m, 	
04.50.Kd, 
11.30.Cp
}
\maketitle



\section{Introduction}

Lorentz symmetry is a well established symmetry in the matter sector and constraints on Lorentz violating phenomenology are quite severe, see for example \cite{arXiv:0708.2889,arXiv:0908.1832,arXiv:0708.1737,arXiv:0805.2548,arXiv:0807.1210,arXiv:0902.1756}. The gravitational sector is far more weakly coupled though, and this allows one to contemplate the idea of Lorentz violating gravity theories. Probably the main motivation for considering such theories, Einstein-aether theory being the typical example \cite{gr-qc/0007031,arXiv:0801.1547}, has so far been the fact that they provide the means for quantifying Lorentz violations in the gravity sector. However, it has been argued recently by Ho\v rava that Lorentz violations in gravity might actually act as a quantum field theory regulator and provide the missing ingredient for an ultraviolet (UV) complete gravity theory \cite{arXiv:0901.3775}. This puts the issue of Lorentz violations in gravity in a different perspective.

The correct UV behavior of the propagators in Ho\v rava's model, which is dubbed Ho\v rava--Lifshitz gravity (see Refs.~\cite{arXiv:1010.3218,Weinfurtner:2010hz,arXiv:1007.5199,arXiv:1103.5587} for brief reviews), hinges on the pre\-sen\-ce of terms in the action which are at least sixth order in spatial derivatives \cite{arXiv:0901.3775} (see also \cite{arXiv:0902.0590,arXiv:0912.4757}). On the other hand, to avoid ghost degrees of freedom one would like to have in the action only terms that are second order in time derivatives. This requires a splitting of spacetime into space and time and leads to Lorentz violations. It is, therefore, perhaps natural to consider such a theory within the framework of an Arnowitt--Deser--Misner decomposition, with $N$ being the lapse, $N_i$ the shift and $g_{ij}$ the induced metric on the spacelike hypersurfaces.
The most general action then is
\beq
S=S_K-S_V,
\eneq
where the kinetic term, which contains all of the time derivatives, is given by
\bea
S_K&=&\frac{2}{k^2}\int{dtd^3x\sqrt{g} N \left(K_{ij}K^{ij}-\lambda K^2\right)}\nonumber\\&=&\frac{2}{k^2}\int dtd^3x\sqrt{g} N K_{ij}G^{ijkl}K_{kl}.
\eea
$K_{ij}$ is the extrinsic curvature of the spacelike hypersurfaces,
\be
K_{ij}=\frac{1}{2N} \left(\dot{g}_{ij}-\nabla_i N_j -\nabla_j N_i\right),
\ee
$\nabla_i$ is the covariant derivative associated with $g_{ij}$, $\lambda$ is a dimensionless coupling and 
$G^{ijkl}$ is the generalized DeWitt metric (the ``metric on the space of metrics").
For the potential term we have
\beq
S_V=\frac{k^2}{8}\int dtd^3x\sqrt{g} N \, V,
\eneq
and $k$ is a coupling of suitable dimensions (we will return to this later). 

To pin down the theory fully one needs to specify $V$. 
The action is not invariant under diffeomorphisms, but the kinetic term is invariant under the subclass of diffeomorphisms that leave the foliation intact, {\em i.e.}~$t\to \tilde{t}(t)$ and $x^i\to \tilde{x}^i(t,x^i)$. Requiring that the potential respects the same symmetry would protect the theory from radiative correction, but at the same time all terms compatible with this symmetry up to at least sixth order in spatial derivatives would have to be taken into account in $V$. This leads to a very large number of terms and an equally large number of independent couplings. 

In order to make the theory more tractable and to limit the proliferation of couplings Ho\v rava imposed two further conditions in Ref.~\cite{arXiv:0901.3775}. The first one is {\em projectability}, that is, the assumption that the lapse is space-independent, $N=N(t)$. This assumption does away with spatial derivatives of $N$ and, combined with invariance under foliation preserving diffeomorphisms, it leads to the requirement $V=V[g_{ij}]$. The second simplicity assumption was {\em detailed balance}, that is, the requirement that V be given by a superpotential $W$ in a specific way. We will discuss this assumption in detail below. For the moment we mention only that the principle of detailed balance is inspired by systems studied in quantum criticality and nonequilibirum critical phenomena and that, combined with projectability, it does indeed limit the number of independent couplings in V to just 3. 

None of the two aforementioned assumptions are strictly necessary. It was argued in Ref.~\cite{arXiv:0904.4464,arXiv:0905.2798} that abandoning detailed balance and considering the most general action with projectability comes at a relatively low cost in additional couplings and does away with some, perhaps unattractive characteristics of the theory with detailed balance, such as a negative (bare) cosmological constant and parity violations. On the other hand, with or without detailed balance, the projectable version of the theory propagates a scalar mode with pathological dynamics and strong coupling at unacceptably low energies \cite{arXiv:0905.2798,Wang:2009yz,Afshordi:2009tt,Charmousis:2009tc,Koyama:2009hc,Blas:2009yd}. The most general action without projectability and without detailed balance manages to avoid all of the aforementioned problems \cite{arXiv:0909.3525}, provided that  there are two fundamental scales in the theory (one related to matter coupling and one to Lorentz breaking) and that these scales are well separated and have the right hierarchy \cite{arXiv:0911.1299,arXiv:0912.0550}. This theory, to which we will be referring as nonprojectable Ho\v rava--Lifshitz gravity, leads to Lorentz violations in the gravity sector at all energies. However, these violations can remain below current experimental accuracy \cite{arXiv:1007.3503}. The potential percolation of Lorentz violations in the matter sector is a much more thorny issue. However, there are mechanisms that can help significantly in this direction, see Ref.~\cite{arXiv:1010.5249} and re\-fe\-ren\-ces therein for a detailed discussion.

An unappealing  characteristic of the full nonprojectable theory is that it has a very large number of independent couplings (order of magnitude $10^2$), which is why restrictions such as detailed balance and pro\-jecta\-bi\-li\-ty were imposed in the first place. At the same time one cannot help but notice that the most disturbing pathologies of the theory with detailed balance and pro\-jecta\-bi\-li\-ty are still present if detailed balance is abandoned without abandoning pro\-jecta\-bi\-li\-ty. This raises the question of whether one could consistently implement detailed balance without pro\-jecta\-bi\-li\-ty and end up with a consistent theory without pathologies and with a much smaller number of independent couplings. This is the question we wish to address in this paper.

In the next section we revisit detailed balance in the projectable version of the theory and provide a critical review of the problems associated with it that have been pointed out in the literature. 
In Sec. \ref{sec:3} we discuss the implementation of detailed balance without pro\-jecta\-bi\-li\-ty and examine whether the resulting theory can be free of pathologies and unappealing characteristics. In Sec. \ref{sec:disc} we summarize our results, discuss their implications, present our conclusions and comment on future perspectives.

Before closing this introduction we would like to mention that the robustness of detailed balance against radiative corrections should not be taken for granted. This is in fact an open issue which has received little attention with or without projectability. This is perhaps justified, as it is a rather complicated problem which might not be worth considering before having a consistent implementation of detailed balance that avoids pathologies at the classical level (similar things can be said for other open questions such as the coupling to matter and the renormalization group flow). In this spirit we focus on the latter question in this paper, hoping to provide a stepping stone for further investigation.


\section{Detailed Balance with projectability}

\subsection{Superpotential and action}
Let us assume that we are working in the projectable version of the theory where $N=N(t)$.
We now impose as an additional symmetry to the theory the so called {\it detailed balance}, which sums up to the requirement that $V$ should be
derivable from a superpotential $W$ as follows:
\beq
\label{Vdef}
V=E^{ij}G_{ijkl}E^{kl},
\eneq
where $E^{ij}$ is given in term of a superpotential $W$ as
\beq
E^{ij}=\frac{1}{\sqrt{g}}\frac{\delta W}{\delta g_{ij}}\,.
\eneq
$G_{ijkl}$ is the inverse of the DeWitt metric which can be written is terms of the induced metric $g_{ij}$ as
\beq
G_{ijkl}=\frac{1}{2}\left(g_{ik}g_{jl}+g_{il}g_{jk}\right)+\frac{\lambda}{1-3\lambda} g_{ij}g_{kl}.
\eneq
The superpotential is supposed to contain all of the possible terms up to a given order in derivatives which are invariant under the symmetry of the theory, {\em i.e.}~invariant under foliation preserving diffeomorphisms. The order in derivatives is dictated by the requirement that the theory be power-counting renormalizable. Minimally this requires sixth order spatial derivatives in the action, which translates to third order spatial derivatives in the superpotential. 

The most general superpotential that satisfies these requirements is
\beq
W=\frac{1}{w^2}\int{\omega_3(\Gamma)}+\mu\int{d^3x\sqrt{g}\left(R-2\Lambda_W\right)}
\eneq 
where  $\omega_3(\Gamma)$ is the gravitational Chern-Simons term and $w$, $\mu$, and $\Lambda_W$ are couplings of suitable dimensions. The potential $V$ corresponding to this superpotential $W$ is
\bea
V&=&\frac{4}{w^4}C_{ij}C^{ij}-\frac{4\mu}{w^2}\epsilon^{ijk}R_{il}\nabla_j R^l_k +\mu^2 R_{ij}R^{ij}\nonumber \\
&&-\frac{\mu^2}{(1-3\lambda)}\left(\frac{1-4\lambda}{4}R^2+\Lambda_W R-3\Lambda_W^2\right)\,.  \label{0.2}
\enea
This is the superpotential used in Ref.~\cite{arXiv:0901.3775}.

\subsection{Known problems and potential solutions}
\label{problems}

The version of the theory with detailed balance is known to be plagued by the following shortcomings:
\begin{enumerate}
\item There is a parity violating term, namely, the term which is fifth order in derivatives. The presence of this term in the action is inevitable if the latter is to contain sixth order derivatives and come from a superpotential as defined above \cite{arXiv:0904.4464,arXiv:0905.2798}.
\item The only sixth order term is the square of the Cotton tensor, which is traceless and vanishes for conformally flat three-dimensional spaces (in this sense, it plays the role of the Weyl tensor in three dimensions, as the latter vanishes identically). As such, it does not contribute to the propagator of the scalar graviton that the theory has. Hence, the scalar mode does not satisfy a sixth order dispersion relation and is not power-counting renormalizable, unlike the spin-2 mode. This spoils the overall UV properties of the theory \cite{arXiv:0901.3775}.
\item The infrared behavior of the scalar mode is plagued by instabilities and strong coupling at unacceptably low energies \cite{Charmousis:2009tc,Blas:2009yd}.
\item The (bare) cosmological constant has the opposite sign from the observed value \cite{arXiv:0904.4464,arXiv:0905.2798}.
\item The (bare) cosmological constant has to be large, much larger than the observed value \cite{arXiv:0907.3121}.
\end{enumerate}
The second problem had been noticed by Ho\v rava already in Ref.~\cite{arXiv:0901.3775} and a resolution has been proposed there. One could add to the superpotential fourth order terms. These would lead to eighth order (super-renormalizable) terms in the action, as well as new sixth and lower order operators. These terms would contribute to the propagator of the scalar mode and power-counting renormalizability would be restored. In fact there are only two fourth order terms one could add to the superpotential, $R^{ij} R_{ij}$ and $R^2$. That is, the improvement in the UV behavior of the scalar graviton comes at a relatively low cost in terms of proliferating the couplings.\footnote{Note that terms with two time and two spatial derivatives of $g_{ij}$, such as $(\nabla_\mu K)(\nabla^\mu K)$ appear to be eighth order operators with $z=3$ scaling but are actually tenth order with $z=4$ scaling.}

Actually, once one has added these terms to cure the behavior of the scalar graviton, a natural resolution to problem 1 emerges: imposing parity invariance explicitly does away with parity violating terms. This would not allow for the presence of $C^{ij}$ in $E^{ij}$, but the renormalizability of the spin-2 graviton is not compromised since there would be both sixth and eighth order terms in $V$. Of course, one might be content with parity violations provided that they come at high enough energies to have remained undetectable so far. If this is the case, provided that the scale of parity violation can be tuned accordingly, problem 1 was not a problem in the first place. 

Problem 3 is not one with an easy remedy. On the other hand, this is not actually a problem specific to detailed balance. In fact, the most general action in Ho\v rava--Lifshitz gravity with projectability and without detailed balance \cite{arXiv:0904.4464,arXiv:0905.2798} exhibits similar problematic behavior when it comes to the infrared dynamics of the scalar graviton  \cite{arXiv:0904.4464,arXiv:0905.2798,Charmousis:2009tc,Blas:2009yd,Wang:2009yz,Afshordi:2009tt,Koyama:2009hc}. Therefore, we consider this problem to be related to projectability and not detailed balance, and we will argue below, in Sec. \ref{remedy}, that it can be addressed successfully in the same manner it has been addressed in the version without detailed balance by doing away with projectability. In other words, we will argue that one needs not abandon detailed balance in order to resolve this issue, but projectability.\footnote{In Refs.~\cite{arXiv:1007.5199,arXiv:1105.0246,arXiv:1109.2609} a different position has been advocated regarding whether the strong coupling of the scalar graviton in the projectable case is truly a problem or an opportunity: it has been claimed that nonperturbative effects, exactly because of the strong coupling, lead to phenomenology very close to that of general relativity via the Vainshtein effect \cite{75796}. At the same time, strong coupling might lead to rapid running of the coupling constant $\lambda$ and act as a remedy for the instability of the scalar mode. Similar statements could be made in the projectable theory with detailed balance. A possible shortcoming in this way of arguing might  be that the arguments used for power-counting renormalizability are essentially based upon the assumption that perturbative treatment does not break down. In Ref.~\cite{arXiv:1109.2609}, it has been claimed that it is enough to solve the momentum constraint nonperturbatively, in which case the power-counting renormalizability arguments are still applicable.}

Lastly, there remain problems 4 and 5 regarding the sign and magnitude of the cosmological constant. We devote the next section to these problems.

\subsection{The size of the cosmological constant}
\label{size}

To get a clearer picture on the various scales involved in the action we perform the following redefinitions of the couplings:
\bea
\label{couplingred}
&&M_{\rm pl}^2=\frac{4}{k^2}, \qquad\qquad M^2_6=\frac{w^2}{2}M^2_{\rm pl}, \nonumber\\&& M_4^2= \frac{M_{\rm pl}^4}{\mu^2},\qquad\qquad \xi=\frac{\Lambda_W}{(1-3\lambda)M_4^2}\,,
\eea
where $M_{\rm pl}$, $M_6$ and $M_4$ have dimensions of a mass, whereas $\xi$ is dimensionless. 
The action corresponding to the potential given in Eq.~(\ref{0.2}) then takes the form
\bea
S_H&=&\frac{M_{\rm pl}^2}{2}\int dt d^3x \sqrt{g}N\Bigg[ K_{ij}K^{ij}-\lambda K^2+\xi R-2\Lambda\nonumber \\
&&\qquad-\frac{1}{M_4^2}R_{ij}R^{ij}+\frac{1-4\lambda}{4 (1-3\lambda)} \frac{1}{M_4^2}R^2\nonumber\\&&\qquad+\frac{2}{M_6^2 M_4}\epsilon^{ijk}R_{il}\nabla_j R^l_k -\frac{1}{M_6^4}C_{ij}C^{ij}\Bigg]\,,  
\eea
where the cosmological constant is
\be
\Lambda=\frac{3}{2}\xi^2(1-3\lambda)M_4^2 \,.
\ee
Clearly general relativity corresponds to $\xi=\lambda=1$ with the terms in the second and third lines being absent. Instead of being parametrized by $k$, $w$, $\mu$ and $\Lambda_W$ the theory is now parametrized by $M_{\rm pl}$, $M_6$,  $M_4$, and $\xi$ (and of course $\lambda$ which is the parameter in the kinetic term in both cases). On the other hand, the cosmological constant $\Lambda$ is {\em not} a free parameter, but instead it is fully determined by the dimensionless parameters $\xi$ and $\lambda$ and the energy scale $M_4$, which is the scale that suppresses one of the fourth order operators in the action.

If we want the theory to be close to general relativity in the infrared, then $\lambda,\xi\sim 1$ to high accuracy. It is already obvious that $\Lambda$ has to be negative in this case, as has been pointed out in the literature ({\em e.g.}~\cite{arXiv:0904.4464,arXiv:0905.2798}). Less attention has been paid to the fact that, what seems to be determining the size of $\Lambda$ is really $M_4$. The latter, as the energy scale that suppresses both of the fourth order operators when $\lambda\sim 1$, will be the energy at which Lorentz-violating effects will become manifest as higher order terms in the dispersion relations. 

There are two classes of observational constraints on Lorentz violations that restrict the size of $M_4$. The first class is purely gravitational constraints. For Lorentz violations to have remained undetected in sub-mm precision tests one would need roughly $M_4\geq 1\div 10{\rm meV}$ (as an optimistic estimate). Much more stringent constraints can be obtained if one considers that Lorentz violations in gravity will percolate the matter sector. If, for instance (and perhaps naively), $M_4$ is taken as a universal scale at which all dispersion relations get modified, it was claimed in Ref.~\cite{arXiv:1007.3503} that the most robust tests of Lorentz violations would require  $M_4\geq 10^{10}$GeV or $M_4\geq 10^{-9} M_{\rm pl}$ (assuming that $M_{\rm pl}$ has its usual value in order for the theory to not differ appreciably from general relativity in the infrared once matter is coupled). This is probably an overoptimistic lower bound, as there are (admittedly less trustworthy) constraints that lead to trans-Planckian values for $M_4$ \cite{arXiv:0708.1737,arXiv:0805.2548,arXiv:0807.1210,arXiv:0902.1756}. However, at the same time it is likely that there are mechanisms that suppress the percolation of Lorentz violations in the matter sector, see Ref.~\cite{arXiv:1010.5249} and references therein. Such mechanisms would actually impose upper bounds on $M_4$, leaving the mild bound coming from gravitational experiments as the only lower bound.

Considering these constraints, we are now ready to re-evaluate the problem with the cosmological constant in the detailed balance scenario. Using only the mildest constraint coming from purely gravitational experiments, the value of the cosmological constant (taking into account the $M_{\rm pl}^2/2$ overall factor in the action) would be (roughly) of the order of $10^{-60}M_{\rm pl}^4$. If more stringent constraints coming from matter are to be imposed this value will get even higher and perhaps larger than $M_{\rm pl}^4$. Consequently, the value of the cosmological constant has to be so large that its negative sign seems to be a secondary problem only: there is at best a 60 orders of magnitude discrepancy between the value required by detailed balance and the observed value. Given that this is a bare cosmological constant, were it allowed to have an arbitrarily small yet negative value, one could just hope for it to be an irrelevant contribution to the total cosmological constant. Note that the vacuum energy problem, or the ``old cosmological constant problem,'' is anyway still an open problem in Ho\v rava gravity, and in a theory which proposes itself as a UV completion of general relativity finding a resolution is pertinent. However, and simply for comparison, the value of the bare cosmological constant when detailed balance is imposed turns out to be at best comparable to the naive estimate of the vacuum energy obtained with $M_{\rm SUSY}$ as a cutoff. 

The fact that the size of the cosmological constant will be related to the size of the energy scale suppressing the fourth order operators has been previously pointed out in Ref.~\cite{arXiv:0907.3121} and it was used there to argue that this leads to a bare cosmological constant that could potentially cancel out the contribution of an equally large vacuum energy. The approach followed there was to provide a heuristic estimate for the vacuum energy and then identify the value that $M_4$ would have to have in order for the aforementioned cancellation to work. This led to near-Planckian values for $M_4$. Here we took an orthogonal approach. We discussed the possible constraints on $M_4$ and derived corresponding constraints on the value of the cosmological constant. Our primary goal was to derive a rough but robust lower limit for the magnitude of the bare cosmological constant. The reason for this is twofold. First of all there is currently no precise and convincing argument that such a cancellation can indeed be achieved without fine tuning. Second, the projectable theory with detailed balance is anyway plagued by problem 3. Therefore, our main concern here was to argue beyond any doubt that the size of the cosmological constant is indeed unacceptably large (the lower bounds we have derived will persist in the non-projectable theory, as we will discuss shortly).

To conclude, the most important problem with the bare cosmological constant in Ho\v rava gravity with detailed balance is not its sign but its magnitude: it has such a large value that, unless one is willing to allow a violation of detailed balance, some sort of self-tuning mechanism along the lines of Ref.~\cite{arXiv:0907.3121} would be the only way to achieve sensible phenomenology.

\section{Detailed balance without projectability}
\label{sec:3}

We have argued above that problems 1 and 2 in the list of Sec. \ref{problems}, {\em i.e.}~parity violations and the UV behavior of the scalar mode, are not real problems, in the sense that they have a straightforward resolution within the framework of projectable Ho\v rava gravity with detailed balance. We have also shown that the main problem with the (bare) cosmological constant is not its sign but its size. Finally, we mentioned already in Sec. \ref{problems} that we consider problem 3, the infrared behavior of the scalar mode, to not be a problem stemming from detailed balance but from projectability and we claimed that it can find a resolution once the latter is abandoned, without having to abandon also the former. We provide support for this claim below.

\subsection{Superpotential and action}

To the best of our knowledge there does not exist in the literature a consistent implementation and consideration of detailed balance without projectability. As has been pointed out in Ref.~\cite{arXiv:0909.3525}, having in mind the version of the theory without detailed balance, once projectability is abandoned one can use, not only the Riemann tensor of $g_{ij}$ and its derivatives, but also the vector 
\beq
a_i=\partial_i \mbox{ln}N\,,
\eneq
in order to construct invariants under foliation preserving diffeomorphisms. In the version of the theory without detailed balance this leads to a proliferation of terms (order $10^2$). In fact all of these terms have to be taken into account as they would anyway be generated by radiative corrections. 

On the other hand, there is, remarkably, only one term one can add to the superpotential $W$ in the version with detailed balance: $a_i a^i$. One then has
\beq
W=\frac{1}{w^2}\int{\omega_3(\Gamma)}+\int d^3x\sqrt{g}\left[\mu\left(R-2\Lambda_W\right)+\beta\,a_i a^i\right]\,, \label{0.1}
\eneq 
where $\beta$ is the new coupling.  This new term has been repeatedly neglected in the literature and abandoning projectability within the framework of detailed balance had been restricted to simply allowing $N$ to have a space dependence without modifying the action. We will show that the presence of this term is crucial when it comes to the low-energy dynamics of the scalar mode (similarly to the version without detailed balance \cite{arXiv:0909.3525}).

The variation of the superpotential with respect to the metric leads to the following additional contribution to $E^{ij}$,
\beq
E^{ij}_{extra}=\frac{1}{\sqrt{g}}\frac{\delta W_{extra}}{\delta g_{ij}}=\beta\left(\frac{1}{2}g^{ij}a_ka^k-a^ia^j\right)\,.
\eneq
Defining the potential in the same way as before (we will return to this subtle issue shortly), according to Eq.~(\ref{Vdef}) one gets the following additional contributions to the action
\bea
S_{extra}&=&\int{dt d^3x \sqrt{g}N \bigg\{\frac{k^2\mu\beta}{4}\bigg[-R_{kl}a^ka^l } \\
&&\qquad+\frac{1-4\lambda}{4(1-3\lambda)}Ra_ka^k+\frac{\Lambda_W}{2(1-3\lambda)}a_ka^k\bigg]\nonumber\\&&\qquad+\frac{k^2\beta}{2w^2}C_{kl}a^ka^l-\frac{k^2\beta^2}{32}\frac{3-8\lambda}{1-3\lambda}(a_ka^k)^2\bigg\}.   \nonumber   \label{0.3}
\enea
Using the coupling redefinitions of Eq.~(\ref{couplingred}) and introducing the extra redefinition
\be
\eta=\frac{\beta\,\xi M_4}{M_{\rm pl}^2}\,,
\ee
leads to the total action
\begin{widetext}
\bea
\label{faction}
S_H&=&\frac{M_{\rm pl}^2}{2}\int dt d^3x \sqrt{g}N\Bigg\{ K_{ij}K^{ij}-\lambda K^2+\xi R-2\Lambda+\eta \,a^ia_i\nonumber \\
&&\quad\qquad-\frac{1}{M_4^2}R_{ij}R^{ij}+\frac{1-4\lambda}{4 (1-3\lambda)} \frac{1}{M_4^2}R^2+\frac{2\eta}{\xi M_4^2}\left[ \frac{1-4\lambda}{4 (1-3\lambda)} R a^ia_i-R_{ij}a^i a^j \right]-\frac{\eta^2}{4\xi^2 M_4^2}\frac{3-8\lambda}{1-3\lambda} (a^ia_i)^2\nonumber\\&&\quad\qquad+\frac{2}{M_6^2 M_4}\epsilon^{ijk}R_{il}\nabla_j R^l_k +\frac{2\eta}{\xi M_6^2 M_4}C^{ij}a_ia_j-\frac{1}{M_6^4}C_{ij}C^{ij}\Bigg\}\,. 
\eea
\end{widetext}
Recovery of Lorentz symmetry would require $\eta\to 0$, as well as $\xi,\lambda \to 1$. The last term in the first line contributes to the low-energy limit of the theory.

Several comments are in order. First of all, the inclusion of the $a^i a_i$ term in the superpotential has no effect in the magnitude of the cosmological constant, so this problem will persist in the theory described by the action in Eq.~(\ref{faction}). Secondly, our implementation of detailed balance in the nonprojectable version of the theory might seem too naive or simplistic. Why not generalize the DeWitt metric further? And why should one stick with Eq.~(\ref{Vdef})? Actually, there do not seem to be any terms one can create which are quadratic in time derivatives of $N$ and are invariant under foliation preserving diffeomorphism. This seems to exclude straightforward generalizations of the DeWitt metric.\footnote{Note also that, in principle one could have new contributions that include one time derivative, such as $K_{ij} a^i a^j$. However, they can be avoided by imposing symmetry under time reversal.} Regarding the generalization of the definition of the potential, indeed a first thing that comes in mind is that $N (\sqrt{-g})^{-1}(\delta W/\delta N)\propto\nabla^2 \ln N$ is an invariant, which could potentially be used to create (higher order) contributions to the action. In absence of a generalized DeWitt metric an unambiguous generalization is, however, not obvious to us.\footnote{In Ref.~\cite{arXiv:1110.5106}, a possible generalization of detailed balanced is proposed (in a version of Ho\v rava--Lifshitz gravity with an extra symmetry). It is not clear to us what motivates the use of $\delta W/\delta a_i$, given that it is actually $N$ that is the fundamental field in the action. On the other hand, in Ref.~\cite{arXiv:0905.3740} generalizations of detailed balance that include the matter fields were considered. We will not follow this approach here.} 

Being left without a guiding principle, and in view of the fact that whether or not (any or which) version of detailed balanced is robust against radiative corrections, we will take the most conservative and simple approach. Note that a generalization of detailed balance would lead to additional terms (and new couplings), but it is unlikely to exclude any of the terms already present in Eq.~(\ref{faction}). So, we will proceed with the action at hand, considering it to be some sort of minimal consistent implementation of detailed balance in nonprojectable Ho\v rava--Lifshitz gravity.

\subsection{Linearization at quadratic order in perturbations}
\label{remedy}

The question that we wish to address next is whether the theory described by the action in Eq.~(\ref{faction}) has improved behavior when it comes to the dynamics of the scalar mode. However, the presence of a large cosmological constant continues to be both a practical complication in this discussion and a phenomenologically undesirable characteristic of the theory under scrutiny. So, perhaps a much better motivated question is the following: if some resolution to the cosmological constant problem were to be found, which would allow one to tune down its magnitude to an acceptable level, would the theory then be free of pathologies when it comes to the dynamics of the scalar? If the answer to this question is positive, then the value of the cosmological constant becomes the only real shortcoming of detailed balance, and indeed this is what we show next.

We, therefore, assume a deus ex machina resolution. We simply set $\Lambda=0$ in Eq.~(\ref{faction}), and ask again if the theory has improved behavior when it comes to the dynamics of the scalar mode. At the low-energy limit, {\em i.e.}~consi\-de\-ring only the second order operators, the answer is obviously yes. This is because, up to this order the theory actually fully coincides with the most general nonprojectable theory considered in Ref.~\cite{arXiv:0909.3525}, which is known to have sensible scalar dynamics at low energies. On the other hand, in the theory with detailed balance that we are consi\-de\-ring here there are significantly less couplings when it comes to higher order terms, and in fact $\xi$ and $\eta$ do enter the coefficients of these terms as well. So, in order to disperse all doubt and show that the scalar field has sensible dynamics at all energies, we will linearize the theory around flat space. 

We start with the total action given in Eq.~(\ref{faction}) and we perturb to quadratic order, considering only scalar perturbations (the theory does not have vector excitations and we are not currently interested in the graviton propagator, whose dynamics were already well behaved in the projectable version of the theory). We, then, have
\beq
N=1+\alpha, \qquad    N_i=\partial_i y, \qquad g_{ij}=e^{2\zeta}\delta_{ij}\,.
\eneq
Our ansatz for the perturbation for $g_{ij}$ differs from the most general scalar perturbation by the term $\partial_i \partial_j E$, but one can use part of the available gauge freedom to set $E=0$. One obtains for the Ricci tensor and the Ricci scalar of $g_{ij}$
\bea
R_{ij}&=&-\partial_i\partial_j\zeta-\delta_{ij}\partial^2\zeta+\partial_i\zeta\partial_j\zeta-\delta_{ij}\partial_k\zeta\partial^k\zeta\,,\\
R&=&-e^{-2\zeta}\left(4\partial^2\zeta+2(\partial\zeta)^2\right)\,,
\eea
where $\partial^2=\delta_{ij}\partial^i\partial^j$.
The quantity $K_{ij}$ appears only quadratically in the action, so we only need to compute it to first order:
\bea
K_{ij}^{(1)}&=&\dot{\zeta}\delta_{ij}-\partial_i \partial_j y \,,\\
K^{(1)}&=&3\dot{\zeta}-\partial^2 y.
\eea
The quadratic action then takes the form
\bea\label{0.4}
S^{(2)}&=&\frac{M_{\rm pl}^2}{2}\int dt d^3x \bigg\{3(1-3\lambda)\dot{\zeta}^2-2(1-3\lambda)\dot{\zeta}(\partial^2 y)\nonumber \\&&+\left(1-\lambda\right)(\partial^2 y)^2 +2\xi(\partial\zeta)^2-4\xi \alpha\partial^2\zeta\nonumber \\
&& +\eta (\partial_i \alpha) (\partial^i \alpha)-\frac{2(1-\lambda)}{1-3\lambda}\frac{1}{M_4^2}(\partial^2\zeta)^2\bigg\}.   
\enea
The $C_{ij} C^{ij}$ term does not contribute because of the conformal properties of the Cotton tensor. The two fifth order operators and the fourth order operators that contain $a_i$ do not contribute as well because they are zero to quadratic order.

Variation with respect to $y$ yields 
\beq
(1-\lambda)\partial^4 y -(1-3\lambda)\partial^2 \dot{\zeta}=0\,,
\ee
which, assuming regularity, leads to
\beq
\partial^2 y=\frac{1-3\lambda}{1-\lambda}\dot{\zeta}\,.     \label{0.9}
\eneq
Variation with respect to $\alpha$ yields
\beq
\eta\partial^2\alpha+2\xi\partial^2\zeta=0\,,
\eneq
which, again imposing regularity, can be solved to give
\beq
\alpha=-\frac{2\xi}{\eta}\zeta.           \label{1.0}
\eneq

We can now use Eqs.~(\ref{0.9}) and (\ref{1.0}) to integrate out the nondynamical fields $y$ and $\alpha$ in favor of the dynamical field $\zeta$. The quadratic action then reads
\bea  \label{1.1}
S^{(2)}\!\!&=&\!\!\frac{M_{\rm pl}^2}{2}\int dt d^3x \bigg\{\frac{2(1-3\lambda)}{1-\lambda}\dot{\zeta}^2+2\xi\left(\frac{2\xi}{\eta}-1\right)\zeta\partial^2\zeta \nonumber  \\&&\qquad\qquad\qquad-\frac{2(1-\lambda)}{1-3\lambda}\frac{1}{M_4^2}(\partial^2\zeta)^2\bigg\}.   
\eea
The dispersion relation for the scalar is then given by
\be
\omega^2=\xi \left(\frac{2\xi}{\eta}-1\right)\frac{1-\lambda}{1-3\lambda}p^2+\frac{1}{M_4^2}\left(\frac{1-\lambda}{1-3\lambda}\right)^2p^4.
\ee
As expected, the low-energy dynamics of the scalar are satisfactory for a significant part of the parameter space (the same part as in the most general nonprojectable theory \cite{arXiv:0909.3525}). In particular, for the scalar to have positive energy (given the sign of the kinetic term for the spin-2 graviton \cite{arXiv:0901.3775,arXiv:0905.2798}) one needs
\beq
\lambda<\frac{1}{3} \qquad {\rm or}  \qquad \lambda>1\,,      \label{2.2}
\eneq
whereas classical stability requires that
\be
\xi\left(\frac{2\xi}{\eta}-1\right)\frac{1-\lambda}{1-3\lambda}>0\,.
\ee
Given the constraints in Eq.~(\ref{2.2}) and the fact that $\xi>0$ for the spin-2 graviton to be stable \cite{arXiv:0901.3775,arXiv:0904.4464,arXiv:0905.2798}, one has
\be
2\xi>\eta>0\,.
\ee

From the coefficient of the $k^4$ term in the dispersion relation one sees directly that any choice for $\lambda$ cannot lead to an instability at higher energies. However, it is also obvious that the scalar satisfies a fourth, and not a sixth, order dispersion relation. This was expected given our discussion about which terms have zero contribution to quadratic order. So, same as in the projectable case, the arguments on which the discussion about the renormalizability properties of the theory is based are compromised, unless we actually go one order higher in the superpotential $W$. Adding fourth order terms in $W$ would lead to both sixth and eight order terms for the scalar, rendering the theory power-counting renormalizable.

The fourth order terms one could add in the superpotential $W$ are
\bea
&& R^2\,, \quad R^{\mu\nu}R_{\mu\nu}\,, \quad R \nabla^i a_i\,, \quad R^{ij}a_i a_j\,,\\
&& R a_i a^i\,,\quad (a_i a^i)^2\,, \quad (\nabla^i a_i)^2\,, \quad a_i a_j \nabla^i a^j\,.\nonumber
\eea
These would add 8 new couplings to the theory.\footnote{Strictly speaking, given that how each new coupling will contribute to the coefficients in the dispersion relation is not obvious once detailed balance has been imposed, one still needs to calculate the full dispersion relations for both the spin-2 and the spin-0 gravitons in order to show without doubt that there is no issue with instabilities at high energies. However, the fairly large number of independent couplings is more than encouraging.} However, after adding these terms one could impose parity invariance without compromising the renormalizability properties of the spin-2 graviton. In this case one would end up with 7 more couplings than the theory in Eq.~(\ref{faction}). In total there would be 12 couplings (not including the coupling to matter). This is roughly an order of magnitude less than the number of couplings in the theory without detailed balance (and up to sixth order operators) \cite{arXiv:0901.3775}. So, even after the addition of the fourth order operators in $W$, detailed balance still provides a significant reduction in the number of couplings.

A discussion about strong coupling is due. As is known, in the theory without detailed balance \cite{arXiv:0901.3775}, the scalar mode appears to be strongly coupled at an energy scale $M_{\rm sc}\leq 10^{-3}\div 10^{-4} M_{\rm pl}$ if only second order operators are considered \cite{arXiv:0911.1299,arXiv:1003.5666,Wang:2010uga}. The strong coupling can be avoided if $M_4<10^{-3}\div 10^{-4} M_{\rm pl}$, {\em i.e.}~if the higher order operators take over at sufficiently low energies \cite{arXiv:0912.0550,arXiv:1007.3503}. The situation will be (qualitatively) similar here, as detailed balance does not restrict the low-energy action but only the higher order operators. 

At this stage it is worth noting also that, as pointed out in Ref.~\cite{arXiv:0911.1299}, apart from the lower bound on $M_4$ coming from strong coupling, there can be an upper bound coming from constraints on modifications in the dispersion relations of the graviton or the matter fields, as discussed also in Sec. \ref{size}. In the former case the bound would be of the order of a few meV so there is no tension. In the latter, if $M_4$ is assumed to be a universal scale at which fourth order correction to the dispersion relations become relevant, the rather optimistic constraints adopted in Ref.~\cite{arXiv:1007.3503} would require $M_4\geq 10^{10}$GeV, comfortably allowing about 6 orders of magnitudes within which $M_4$ could lie. However, to the extent at which the ultra high energy constraints obtained in \cite{arXiv:0708.1737,arXiv:0805.2548,arXiv:0807.1210,arXiv:0902.1756} can be trusted, $M_4$ has to be trans-Planckian, and there is no window of opportunity if the strong coupling constraint is also taken into account. In this case, a mechanism that would control the percolation of Lorentz violations in the matter sector, such as the one proposed in Ref.~\cite{arXiv:1010.5249}, is necessary. 

\section{Discussion}
\label{sec:disc}

We have revisited the idea of detailed balance in Ho\v rava--Lifshitz gravity, as a way to restrict the proliferation of independent couplings. We first considered the projectable version of the theory, in which this principle had been initially implemented. We listed the various shortcomings usually associated with detailed balanced and discussed some potential resolutions that have been proposed for some of them. The problems that cannot find a resolution within the framework of projectability and detailed balance were the sign and magnitude of the bare cosmological constant and the dynamical inconsistencies associated with the scalar mode. 

We have shown that the latter of the two problems is actually related to projectability and not detailed balance. That is, we have shown that a nonprojectable formulation of the theory with detailed balance would lead to sensible dynamics for the scalar mode and the same low-energy phenomenology as the version without detailed balance, were the magnitude problem of the cosmological constant to find a resolution. The theory would be required to have fourth order derivative terms in the superpotential, but it would still have a number of independent couplings which would be roughly an order of magnitude lower than the  version without detailed ba\-lance.

Could the magnitude and the sign of the bare cosmological constant, which are the only shortcomings that persist once projectability is abandoned, be blessings in disguise, as proposed in Ref.~\cite{arXiv:0907.3121} for the projectable case? That is, could the bare cosmological constant end up canceling out the contribution of the vacuum energy, leaving behind a tiny residual that would account for the observed value? Certainly at this stage, and with the current poor level of understanding of the vacuum energy problem in Ho\v rava--Lifshitz gravity, such a statement is at the level of wishful thinking. Nevertheless, one cannot exclude the possibility. This provided some extra motivation to consider the vacuum energy problem, which is anyway a pressing problem for a theory that aspires to be a UV completion of general relativity.

In absence of a miraculous cancellation between bare cosmological constant and vacuum energy, one could consider the idea that the bare cosmological constant could be tuned down by a soft breaking of detailed balance, which raises the question whether such a soft breaking would not affect higher order terms as well. In fact, this brings one back to the key issue of whether detailed balance can anyway be robust against radiative corrections. This question becomes much more interesting in the light of the fact that a dynamically consistent theory that satisfies detailed balance does indeed exist.

\begin{acknowledgments}
TPS would like to thank Ted Jacobson for stimulating questions that led to the initiation of this project. We are grateful to Petr Ho\v rava, Matt Visser, and Silke Weinfurtner for enlightening discussions. 
\end{acknowledgments}

\end{document}